\documentclass{PoS}

\title{Cosmic Rays from Heavy Dark Matter from the Galactic Center}

\ShortTitle{Cosmic Rays from Heavy Dark Matter from the Galactic Center}

\author{\speaker{Jose A. R. CEMBRANOS}
         \\
        Departamento de  F\'{\i}sica Te\'orica I, Universidad Complutense de Madrid, E-28040 Madrid, Spain\\
        E-mail: \email{cembra@fis.ucm.es}}

\author{Viviana GAMMALDI\\
        Departamento de  F\'{\i}sica Te\'orica I, Universidad Complutense de Madrid, E-28040 Madrid, Spain\\
        E-mail: \email{vivigamm@pas.ucm.es}}

\author{Antonio L. MAROTO\\
        Departamento de  F\'{\i}sica Te\'orica I, Universidad Complutense de Madrid, E-28040 Madrid, Spain\\
        E-mail: \email{maroto@fis.ucm.es}}

\abstract{
The gamma-ray fluxes observed by the High Energy Stereoscopic System (HESS) from the J1745-290 Galactic Center source is well fitted by the secondary photons coming from Dark Matter (DM) annihilation in particle-antiparticle standard model pairs over a diffuse power-law background. The spectral features of the signal are consistent with different channels: light quarks, electro-weak gauge bosons and top-antitop production. The amount of photons and morphology of the signal localized within a region of few parsecs, require compressed DM profiles as those resulting from baryonic contraction, which offer large enhancements in the signal over DM alone simulations. The fits return a heavy WIMP, with a mass above 10 TeV, but well below the unitarity limit for thermal relic annihilation.
The fitted background spectral index is compatible with the Fermi-Large Area Telescope (LAT) data from the same region. This possibility can be potentially tested with the observations of other high energy cosmic rays.
}

\FullConference{The European Physical Society Conference on High Energy Physics \\
                 18-24 July, 2013\\
                 Stockholm, Sweden}

\begin{document}

\section{Gamma rays from the galactic center and the dark matter hypothesis}

We have studied the possibility of explaining the gamma ray data \cite{HESS} observed by the High Energy Stereoscopic System (HESS) from the central part of
our galaxy by being partially produced by Dark Matter (DM) annihilations or decays \cite{Cembranos:2012nj,Belikov:2012ty}.
The complexity of the region and the amount of
delocalized emitting sources justifies the hypothesis of a non-negligible background. DM annihilations or decays into single standard model
particle-antiparticle channels provide good fits if the DM signal is complemented with such a background, that it is consistent with Fermi-LAT
measurements \cite{Fermi}.

The fits return a DM mass between $15$ and $110 \; \rm{TeV}$ depending on the studied channel \cite{Ce10}. Leptonic channels
are clearly disfavored, but hadronic channels such us the $d\bar d$, or electroweak channels such as the $W^+W^-$ and $ZZ$ channels,
offer very good consistency with $\chi^2/dof=0.73$, $0.84$ and $0.85$ respectively \cite{Cembranos:2012nj}.
The morphology of the signal requires compressed dark halos
as the ones that take into account baryonic dissipation \cite{Blumenthal}.

A motivated DM candidate which could have high enough masses and account for the right abundance of DM as a thermal relic,
is the branon field \cite{branons,branonsgamma}, which corresponds to brane fluctuations in flexible brane-world models.
We have proved that a branon mass of $M \simeq 50.6$ TeV provides a good fit to the mentioned HESS data \cite{Cembranos:2012nj}.
%At these masses, its main contribution to gamma rays comes from the annihilation into gauge bosons.

\section{Conclusions and future work}

The heavy DM masses required for fitting the HESS data are practically unconstrained by direct detection searches
or colliders experiments \cite{lab}.
On the contrary, this possibility can be tested with the observations of other cosmic-rays \cite{cosmics} from the GC and from other astrophysical objects.
For example, high energy neutrinos could confirm this hypothesis at IceCube or ANTARES for angular resolutions below one degree.

\acknowledgments{
This work has been supported by MICINN (Spain) project numbers FIS2011-23000, FPA2011-27853-01 and Consolider-Ingenio MULTIDARK CSD2009-00064.
}

\end{document}